\documentclass[aps,prd,reprint,superscriptaddress,preprintnumbers,longbibliography]{revtex4-1}

\usepackage{amsmath,amssymb,bm,mathrsfs,bbold,dsfont}
\usepackage{xfrac}
\usepackage[dvipsnames]{xcolor}
\usepackage{physics}
\usepackage{slashed}
\usepackage{multirow}
\usepackage{booktabs}

\usepackage{epsfig}
\usepackage{psfrag}
\usepackage{tikz}
\usepackage{tikz-feynman}
\usepackage[caption=false]{subfig}  
\PassOptionsToPackage{caption=false}{subfig}

\usepackage[sort&compress]{natbib}
\usepackage[colorlinks=true,
    linkcolor=red,
    citecolor=blue,
    urlcolor=blue,
    filecolor=blue,
    anchorcolor=blue,
    menucolor=blue,
    linktocpage=true,
    pdfproducer=medialab,
    pdfa=true
]{hyperref}
\usepackage{cleveref}

\usepackage{setspace}
\usepackage{relsize}
\usepackage{enumerate}
\usepackage{comment}
\usepackage[normalem]{ulem}
\usepackage{srcltx}
\usepackage{xcolor}
\usepackage{diagbox}
\usepackage{appendix}

\newcommand{\keV}{\,\mathrm{keV}}
\newcommand{\GeV}{\,\mathrm{GeV}}
\newcommand{\TeV}{\,\mathrm{TeV}}
\newcommand{\kms}{\,\mathrm{km/s}}
\newcommand{\TRH}{T_{\rm RH}}
\newcommand{\Mpl}{M_{\rm Pl}}
\newcommand{\vmin}{v_{\rm min}}
\newcommand{\vesc}{v_{\rm esc}}

\begin{document}

\preprint{ MIT-CTP/6106}

\newcommand{\ourtitle}{\boldmath Heavy Higgsino Interpretation of the LZ Event}

\title{\ourtitle}

\author{Kevin Langhoff}
\email{langhoff@mit.edu}
\affiliation{Center for Theoretical Physics -- a Leinweber Institute, Massachusetts Institute of Technology, Cambridge, MA 02139, USA}

\begin{abstract}\noindent
The LZ experiment has recently reported a $248\keV$ nuclear recoil event in a region where backgrounds are expected to be low. A pure higgsino scattering inelastically through a $Z$-boson is an exciting possible interpretation of this event, but at the mass required for thermal freeze-out to account for all dark matter, $m_\chi = 1.1\TeV$, it predicts several events in the empty high-energy sideband corresponding to $E_R\gtrsim350\keV$ and is further excluded by searches for high energy neutrinos coming from the Sun by the IceCube detector. We fit the higgsino mass scale and neutral state mass splitting, $m_\chi$ and $\delta$, to the event and the empty sideband. We additionally recompute solar capture bounds as a function of $m_\chi$. The LZ data alone are best fit at low mass, $m_\chi \approx (240,\,450)\GeV$, but is excluded by solar capture. The parameter space most consistent with the event, the empty sideband and the solar capture bound is $m_\chi \approx (10^5,\,10^6)\GeV$ with $\delta \approx (330,\,480)\keV$, with some dependence on DM halo modeling. Such a higgsino must be produced by some other method than standard freeze-out; this can be achieved by freezing-out during a period of early matter domination which ends by reheating the universe to temperature $T_{\rm RH}\approx 1\TeV$. We give a simple proof of concept model which does this via a Kim-Nilles mechanism and therefore connects to the origin of the Higgsino mass scale and solves the strong CP problem.
\end{abstract}

\maketitle


\section{Introduction}

Recently, the LUX-ZEPLIN (LZ) collaboration reported the observation of a single event with nuclear recoil energy of $E_R = 248 \pm 23({\rm stat}) \pm 23({\rm sys}){~\rm keV}$~\cite{LZ:2026axp}. A tantalizing possibility is that this event could have arisen from a collision with pure higgsino dark matter (DM)~\cite{Fan:2026kxx,Freese:2026sga,Wu:2026nhi,DiMauro:2026ldr}. The pure \emph{thermal higgsino} mass eigenstates are fixed near $m_\chi = 1.1$ TeV under the assumptions that it makes up all of DM and is produced by thermal freeze-out~\cite{Bottaro:2022one}. To account for this event, the mass splitting between neutral state must be $\delta \sim (350,500)~{\rm keV}$\cite{Fan:2026kxx}. In the MSSM with electroweakinos hierarchically heavier than the higgsino, this mass splitting is given by
\begin{align}
    \delta \approx m_Z^2\left(\frac{s_W^2}{M_1} + \frac{c_W^2}{M_2}\right)
\end{align}
where $M_1$ and $M_2$ are bino and wino masses respectively; this would suggest a large hierarchy between the higgsino mass at the TeV scale and gaugino masses at $\mathcal{O}(10\,{\rm PeV})$.

However, there are some challenges to this proposal. First, this thermal higgsino interpretation would suggest the existence of additional nuclear recoil events with even higher recoil energies which are not observed~\cite{Rodd:2026tyn}. Second, such higgsinos would be captured in the core of the Sun and annihilate to high-energy neutrinos at a rate excluded by IceCube unless $\delta \gtrsim 500~\keV$~\cite{Pospelov:2026ewn,IceCube:2025fcu}.

In this Letter, we show that the higgsino DM interpretation in least tension with the 248~keV nuclear recoil event and the above two observations is that the event arose from the collision with a heavy pure higgsino dark matter particle with mass $m_\chi \sim (10^5,\, 10^6)~{\rm GeV}$. Such a heavy higgsino can not be produced through standard thermal freeze-out and requires modifications of early universe cosmology from pure radiation domination. Making the higgsino heavy does have the added benefit of reducing tuning as 1-loop corrections from heavy Higgs and gauginos must be tuned at the 10\% level to maintain the hierarchy in mass scales~\cite{Fan:2026kxx}.

In Sec.~\ref{sec:LZ} we calculate a 2D likelihood in the pure higgsino parameter space $(m_\chi,\,\delta)$ using the $248\keV$ event and the absence of event in the high-energy sideband. In Sec.~\ref{sec:Solar} we calculate the solar capture bounds of \cite{Pospelov:2026ewn} at all masses and show how these bounds weaken for heavy higgsinos. In Sec.~\ref{Sec:Origin} we present a proof of concept model which allows a cosmology where such heavy higgsinos are all of DM and connects this to the origin of the MSSM parameter $\mu$ determine the higgsino mass through a Kim-Nilles term~\cite{Kim:1983dt}. Finally, we conclude in Sec.~\ref{sec:Conclusion}.

\begin{figure*}[t!]
\centering
\includegraphics[width=\textwidth]{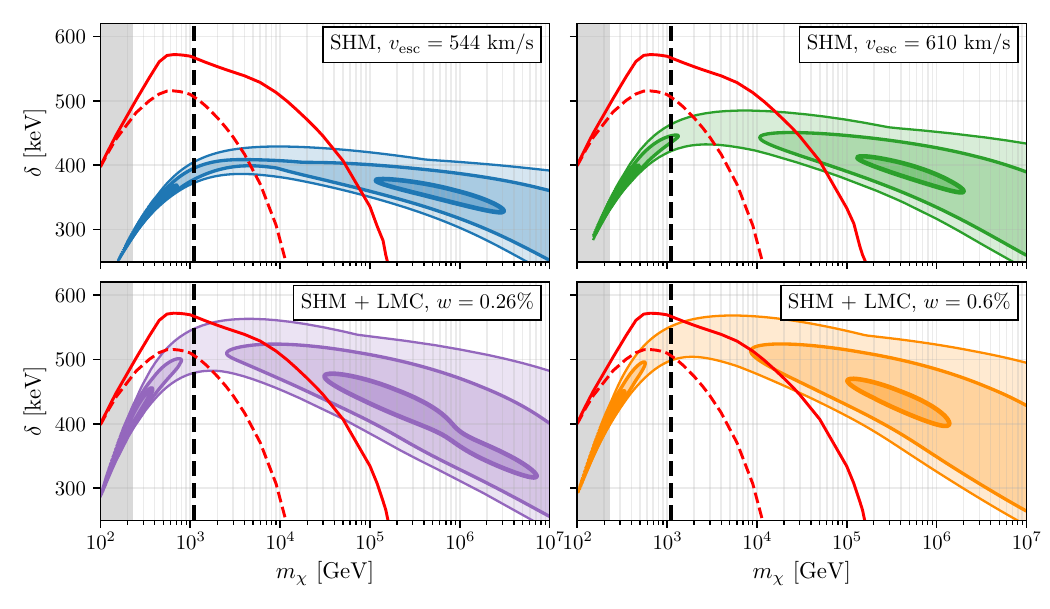}
\caption{Regions of $(m_\chi,\delta)$ pure higgsino parameter space where fitting to the LZ event and an empty high-energy sideband gives $\Delta\chi^2 < 2.30$, $6.18$ and $11.83$ for four halo models. Red curves are lower bounds on $\delta$ from solar capture, computed following Ref.~\cite{Pospelov:2026ewn}, as a function of $m_\chi$ with (and without) thermalization via loop-induced elastic scattering shown as a solid (dashed) curve. The black dashed line is the thermal higgsino at $1.1\TeV$, the grey band is the LHC lower bound of $225\GeV$ from disappearing tracks~\cite{ATLAS:2026hnb}. }
\label{fig:contours}
\end{figure*}

\section{Pure Higgsino Interpretation of LZ Event} \label{sec:LZ}

We first consider the interpretation of LZ results as coming from higgsino scattering following and perform a 2-dimensional likelihood fit over the pure higgsino parameter space $(m_\chi,\,\delta)$ over the observation of a single event at $E_R = 248 \pm 23({\rm stat}) \pm 23({\rm sys}){~\rm keV}$ in the LZ region of interest (ROI), with $E_R<270~{\rm keV}$, and the non-observation of any event in the high-energy sideband, defined by $800<S1_c<1700$~phd and $10^{2.75}<S2_c<10^{4.3}$~phd, which corresponds to $350<E_R<675\keV$ for nuclear recoils~\cite{LZ:2026axp,Rodd:2026tyn}.

The two neutral higgsino mass eigenstates with mass splitting $\delta$ couple to the $Z$ boson off-diagonally such that tree level scattering with nuclei is purely inelastic. The differential nuclear recoil rate for these inelastic collisions is
\begin{align} \label{eq:rate}
    \frac{dR}{dE_R} = \frac{\rho_\chi \sigma^0_{A}Q_W^2}{2m_\chi \,\mu_A^2}\,F^2(E_R)\, \eta(\vmin(E_R)),
\end{align}
where $\rho_\chi$ is the local DM density (which we take to be $0.3\,\GeV/{\rm cm}^3$), $\sigma_A^0 = G_F^2\mu_A^2/2\pi$ is the $Z$-exchange cross section at zero momentum exchange~\cite{Essig:2007az,Rodd:2026tyn}, $\mu_A$ is the reduced mass, $F$ the Helm form factor~\cite{Helm:1956zz,Lewin:1995rx}, $Q_W = N - (1-4s_W^2)Z$, and $\eta(\vmin) = \int_{\vmin} d^3v\, f(\bm v)/v$ in the laboratory frame, averaged over a year~\cite{McCabe:2010zh}, and 
\begin{align} \label{eq:vmin}
\vmin(E_R) &= \sqrt{\frac{\delta}{2\mu_A}}\left(\sqrt{\frac{E_R}{E_R^*}} + \sqrt{\frac{E_R^*}{E_R}}\right),
\end{align}
with $E_R^* = (\mu_A/m_A)\,\delta$.

Note that the threshold speed $\sqrt{2\delta/\mu_A}$ is reached only at $E_R = E_R^*$. Since $\eta(\vmin)$ is a steeply falling function of $\vmin$ (especially at the high velocities which will be of interest to us), this tends to localize recoil energies around $E_R^*$, with additional effects coming from the form factor. This suggests that in order to see an event with $E_R \approx 248\,\keV$ while not seeing any event in the high-energy side band and avoid the tension pointed out in \cite{Rodd:2026tyn}, it is beneficial to reduce $E_R^*$ either by reducing $\mu_A$ (i.e. bring $m_\chi$ closer to $m_A$) or by decreasing $\delta$.

The phase space factor, $\eta(\vmin)$, depends strongly on the modeling of the tail of the velocity distribution ~\cite{Fan:2026kxx}. Following Ref.~\cite{Rodd:2026tyn}, we model the distribution by
\begin{align}
&f(\bm v) = (1-w)\,f_{\rm SHM}(\bm v) + w\,f_{\rm LMC}(\bm v),\\
&f_{\rm SHM}(\bm v) \propto e^{-v^2/v_0^2}\,\Theta(\vesc - v), \\
&f_{\rm LMC}(\bm v) \propto e^{-|\bm v - \bm v_b|^2/\sigma_b^2}\,\Theta(v_{\rm cut} - |\bm v - \bm v_b|),
\end{align}
 and fix the overall normalization using $\rho_\chi$. For the standard halo model (SHM) we use $v_0 = 238~\kms$ and consider two different escape velocities; $\vesc = 544$ and $610~\kms$. The second term is a high-velocity component from the Large Magellanic Cloud (LMC)~\cite{Besla:2019xbx}; following Ref.~\cite{Rodd:2026tyn}, we use a Gaussian of width $\sigma_b = 100~\kms$, $|\bm v_b| = 570~\kms$ oriented at $\cos\beta = -0.71$ to the Sun's velocity, $v_{\rm cut} = 200~\kms$, and consider mass fractions $w = 0.26\%$ and $0.6\%$.

For detector modeling, we use digitized efficiencies from the LZ analysis~\cite{LZ:2026axp} and assume perfect efficiency for the high energy sideband as a benchmark. We use an approximate signal-only likelihood for an isolated event in the ROI and the empty high-energy sideband. Defining $N_{\rm ROI}$ and $N_{\rm SB}$ as the expected number of events in the ROI and high-energy sideband respectively, and $p(E_R)$ as the expected event density obtained from folding detector efficiency and recoil energy resolution into Eq.~\eqref{eq:rate}, we obtain a likelihood
\begin{align}
    -\log\mathcal L = N_{\rm ROI} + N_{\rm SB} - \log p(E_R = 248~\keV).
\end{align}
We measure the likelihood difference relative to the global best fit within the higgsino model, $\Delta\chi^2 = 2\,\Delta(-\log\mathcal L)$.  

In Fig.~\ref{fig:contours} we show likelihood contours with $\Delta\chi^2 < 2.30$, $6.18$ and $11.83$ for the two parameter fit used as a \emph{heuristic guide} to the relative quality of fit within this model. The global best fit occurs at low masses, i.e. $m_\chi = (200,\,400)\GeV$ due to the decrease in $E_R^*$ from the decrease in $\mu_A$ which correlates with a decrease in $N_{\rm SB}$. At high masses, the DM density falls as $1/m_\chi$ such that required collision rate can only be recovered if the phase space contributions grow by decreasing $\delta$ (and therefore $\vmin$). Since for these masses $\mu_A \approx m_A$, $E_R^*$ is determined entirely by $\delta$, and the local minimum occurs when this is optimal for the given Halo model. We also see that there is a tension in fitting the data with the thermal higgsino at $m_\chi = 1.1\TeV$ in agreement with~\cite{Rodd:2026tyn}. We show the best fit higgsino parameters for the four different halo models and different higgsino mass assumptions in Tab.~\ref{tab:fit} as well as the expected number of events in the ROI and high-energy sideband (i.e. $N_{\rm ROI}$ and $N_{\rm SB}$). While the heavy higgsino best fit point lies within the $\Delta\chi^2<2.30$ contour, it does not fit the data quite as well. This is because for low masses one can tune $m_\chi$ and $\delta$ to change the rate and $E_R^*$ independently. At high masses this is no longer true since $\mu_A\approx m_A$.

One prediction of this fit is that the heavy higgsino best fit parameters give comparable \emph{expected} (potentially fractional) number of events in the ROI and the sideband, $N_{\rm SB}/N_{\rm ROI} \approx 0.8$ to $1$, whereas the light branch gives a ratio of $\approx 0.1$ to $0.2$ and the thermal higgsino $\gtrsim 15$.

\begin{table}[h!]
\caption{Best fit points for different halo models and higgsino mass assumptions; the light global minimum, the thermal higgsino ($m_\chi = 1.1~\TeV$) and the heavy local minimum ($\Delta\chi^2$ is nearly flat in $m_\chi$ for the heavy local minimum, so the tabulated mass is representative). Halo models are the SHM with $\vesc = 544$ or $610~\kms$, and the SHM ($\vesc = 544~\kms$) with an LMC fraction $w = 0.26\%$ or $0.6\%$. $N_{\rm ROI}$ and $N_{\rm SB}$ are the expected counts in the ROI and the high-energy sideband.}
\label{tab:fit}
\renewcommand{\arraystretch}{1.2}
\begin{ruledtabular}
\begin{tabular}{ll|cccccc}
 & & $m_\chi$ [GeV] & $\delta$ [keV] & $N_{\rm ROI}$ & $N_{\rm SB}$ & $\Delta\chi^2$ \\
\hline
\multirow{3}{*}{SHM 544}
 & light   & $4.50\times10^{2}$ & 344 & 0.88 & 0.10 & 0    \\
 & thermal & $1.10\times10^{3}$ & 384 & 0.13 & 2.13 & 5.5  \\
 & heavy   & $3.20\times10^{5}$ & 365 & 0.55 & 0.47 & 1.9  \\
\hline
\multirow{3}{*}{SHM 610}
 & light   & $3.33\times10^{2}$ & 371 & 0.90 & 0.11 & 0    \\
 & thermal & $1.10\times10^{3}$ & 437 & 0.05 & 3.16 & 9.3  \\
 & heavy   & $2.72\times10^{5}$ & 401 & 0.51 & 0.49 & 2.1  \\
\hline
\multirow{3}{*}{LMC 0.26}
 & light   & $2.46\times10^{2}$ & 414 & 0.77 & 0.13 & 0    \\
 & thermal & $1.10\times10^{3}$ & 504 & 0.10 & 2.77 & 6.8  \\
 & heavy   & $1.02\times10^{5}$ & 449 & 0.54 & 0.45 & 1.8  \\
\hline
\multirow{3}{*}{LMC 0.6}
 & light   & $2.39\times10^{2}$ & 415 & 0.82 & 0.12 & 0    \\
 & thermal & $1.10\times10^{3}$ & 516 & 0.03 & 3.77 & 11 \\
 & heavy   & $2.50\times10^{5}$ & 447 & 0.56 & 0.44 & 2.1  \\
\end{tabular}
\end{ruledtabular}
\end{table}

\section{Solar Capture vs Mass} \label{sec:Solar}

In \cite{Pospelov:2026ewn}, bounds on $\delta$ for higgsino dark matter were given for the thermal higgsino mass of $1.1\TeV$. In this section we discuss how these results extend to different masses. Specifically, these bounds weaken for heavier higgsinos since the kinetic energy lost in a single inelastic scattering event of $\mathcal{O}(\delta)$ becomes insignificant and very rarely causes the higgsino to get trapped. For low mass higgsinos, solar capture bounds on $\delta$ also weaken because the collision speed required to excite the higgsino grows as the reduced mass decreases.

The same inelastic scattering which could have produced the LZ event also captures dark matter inside the Sun. A Higgsino with speed $v$ far from the Sun arrives to a distance $r$ from the center of the Sun with a speed $w = \sqrt{v^2 + v_{\rm esc}(r)^2}$ and is captured if it has speed less than $v_{\rm esc}(r)$ after a collision. This occurs for nuclear recoil energies $E_R\geq \frac{1}{2}m_\chi v^2 - \delta$.

\begin{figure}[t]
    \centering
    \includegraphics[width=1\linewidth]{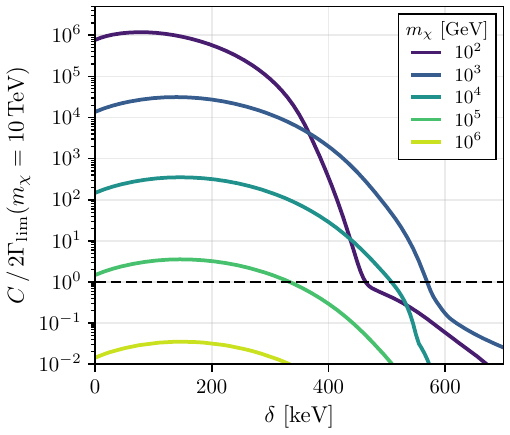}
    \caption{Solar capture rate normalized to the IceCube annihilation rate upper bound for $m_\chi = 10\TeV$ as a function of $\delta$ for several values of $m_\chi$. While $\Gamma_{\rm lim}(m_\chi)$ has strong $m_\chi$ dependence (for example bounds on the annihilation rate at $m_\chi = 100\GeV$ are 400 times weaker than at $10 TeV$) for $m_\chi\lesssim 1 TeV$, this limit is relatively constant such that the lower bound on delta can be approximated by eye as the intersection of the curve with the dashed line. }
    \label{fig:Capture_vs_Delta}
\end{figure}

The capture rate is~\cite{Pospelov:2026ewn,Nussinov:2009ft,Menon:2009qj}
\begin{align}\label{eq:capture}
    C = &\sum_A \frac{\rho_\chi\,\sigma_A^0 Q_W^2}{2 m_\chi\,\mu_A^2}
\int dV\, \rho_A(r) \int dv\, \frac{f(v)}{v} \times \notag \\
&\int_{E_{\min}(v)}^{E_{\max}(v)} dE_R\, F^2(E_R),
\end{align}
where $f_\odot(v)$ is the dark matter \textit{speed} distribution in the Sun's frame far from the Sun, $\rho_A(r)$ is the mass density of element $A$ at radius $r$~\cite{Bahcall:2004pz}, and $E_{\min}(v)$ and $E_{\max}(v)$ delimit the recoil energies that are both kinematically allowed and large enough to leave the higgsino bound to the Sun. The kinematic limits on all possible recoil energies are $E_\pm=\frac{\mu_A^2}{2m_A}(w\pm w')^2$ with ${w'}^2=w^2-2\delta/\mu_A$ such that the limits of integration for the recoil energy in the capture rate are $E_{\min}=\max\left(E_-,\frac{1}{2} m_\chi v^2-\delta\right)$ and $E_{\max}=E_+$. Importantly, the dilution in the dark matter number density as $1/m_\chi$ and and additional factor of $1/m_\chi$ from shrinking phase space for the higgsino to be captured cause the capture rate to shrink as $1/m_\chi^2$ for $m_\chi\gtrsim 1\TeV$, see Fig.~\ref{fig:Capture_vs_Delta}.

The higgsinos captured by the Sun can annihilate to weak bosons and subsequently decay to neutrinos and be observed by IceCube~\cite{IceCube:2025fcu} which sets bounds on the annihilation rate. Sufficiently energetic neutrinos produced by annihilation in the solar core undergo absorption, neutral-current scattering, and regeneration through tau production and decay which reduce the dependence of the neutrino spectrum shape on the dark matter mass for very large masses~\cite{Cirelli:2005gh}. This is seen by weak mass dependence of the inferred $W^+W^-$ annihilation-rate limit near $m_\chi=10\TeV$~\cite{Pospelov:2026ewn}. This motivates
\begin{align}
    \Gamma_{\rm lim}(m_\chi)
    \approx \Gamma_{\rm lim}^{WW}(10\,{\rm TeV}),
    \qquad m_\chi>10\,{\rm TeV}.
\end{align}
We use this extrapolation as a benchmark for solar capture constraints for $m_\chi>10\TeV$.

The annihilation rate depends on the time to reach equilibrium between capture and annihilation, $\tau_{\rm eq}$, as $\Gamma = \tfrac12 C\tanh^2(t_\odot/\tau_{\rm eq})$ (where $t_\odot$ is the age of the Sun). Whether equilibrium is reached depends on how concentrated the captured population is. 

Inelastic collisions stop once the higgsino is no longer energetic enough to up scatter to the higher mass state. Loop-induced elastic scattering can further cool the captured higgsinos~\cite{Hisano:2011cs,Pospelov:2026ewn}. In the limit where annihilation is in equilibrium with capture we have $\Gamma = C/2$ which gives the solid red curves in Fig.~\ref{fig:contours}. The red dashed curve is obtained under the assumption elastic scattering and Sommerfeld enhancement are ignored. The two red curves therefore represent limiting behavior. In \cite{Pospelov:2026ewn}, it is shown that the solid line should represent the bound at $1.1\TeV$; a careful analysis including Sommerfeld enhancement and 1-loop elastic scattering would be required to identify where the true bound exists for different masses; however, the limiting behavior is sufficient to justify our main claims. Bounds on $\delta$ change by less than a few $\keV$ between the halo models, since capture is not very sensitive to the high-velocity tail.

Compared with the LZ regions of Fig.~\ref{fig:contours}, the low mass higgsino parameter space preferred by the LZ results lies well below the bound for every halo model. The best fit values of $\delta$ for the thermal higgsino are similarly excluded. An LMC tail raises the thermal higgsino splitting to within $25~\keV$ of the tree-level bound, but the same tail fills the high-energy sideband, so the two constraints are complementary. If higgsino DM does explain the LZ event, this suggests it must be very heavy, i.e. $m_\chi \approx (10^5,\,10^6)~\GeV$.

\section{The Origin of the Heavy Higgsino} \label{Sec:Origin}
If a higgsino with $m_\chi \approx (10^5,\,10^6)~\GeV$ undergoes standard freeze-out, it would greatly overproduce dark matter since $\Omega_\chi \propto 1/\expval{\sigma v}\propto m_\chi^2$. Therefore, its production method must be modified to dilute this abundance. In this section, we give one simple proof of concept as to how this can be done. We motivate the specific proof of concept model by connecting to the origin of the MSSM $\mu$ parameter and a solution to the strong-CP problem using the Kim-Nilles mechanism \cite{Kim:1983dt}, although the general mechanism we give for diluting the higgsino abundance applies more generally~\cite{Giudice:2000ex}.

In order to dilute the abundance of a heavy higgsino, we consider a cosmology with a period of early matter domination (EMD) of a scalar $X$ which reheats the universe at temperature $T_{\rm RH}\approx \sqrt{\Mpl \Gamma_X}$. During this EMD era, decays of $X$ populates a subdominant bath of thermal radiation which cools with the expansion of the universe at $T \propto a^{-3/8}$. The higgsino freezes out of this thermal bath at temperature $T_f$ (which we assume happens during the EMD era although could also happen just before) and its yield further dilutes as $(T/T_f)^5$ until reheating by the entropy released in $X$ decays. The yield at reheating is \cite{Giudice:2000ex}
\begin{align}
    Y_\chi \approx \frac{H(T_f)}{\expval{\sigma v} s(T_f)}\left(\frac{T_{\rm RH}}{T_f}\right)^5 \approx \frac{\kappa \, T_{\rm RH}^3}{\Mpl \expval{\sigma v} T_f^4}
\end{align}
where numerically it is found that $\kappa \approx 0.02$ and $m_\chi/T_f \approx 13$. Relative to standard thermal freeze-out, this reduces the higgsino abundance proportional to $(\TRH/T_f)^3$ such that obtaining the correct abundance requires 
\begin{align}
    \TRH \approx 1.5\TeV\, \left(\frac{m_\chi}{10^6\GeV}\right)^{1/3}\left(\frac{x_f}{13}\right)^{-4/3}
\end{align}
where $x_f$ has only slow logarithmic dependence.

The above is very general, however, we must also require either $m_X < 2 m_\chi$ in order to prevent decays into R-parity odd states or tune the branching ratio into R-parity odd states to be small. For example, if $X$ decayed into R-parity odd states with branching ratio $\rm{BR}$, the higgsino yield by number conservation would become $Y_\chi \approx 3 \rm{BR} \TRH/2m_X$ which would require  $\rm{BR}\lesssim 10^{-15}$ or so to get the correct relic abundance. Instead of considering how to obtain such a large suppression we will focus on the scenario $m_X < 2 m_\chi$.

This upper bound on the decaying particle mass slightly complicates the story as an compelling candidate to dilute the abundance would be a long lived modulus coupling with gravitational strength and therefore decaying with rate $\Gamma_X \sim m_X^3/\Mpl^2$; however, requiring $m_X <2m_\chi \lesssim 10^6\GeV$ (from the above fit to LZ data) implies $\TRH \lesssim 1\GeV$ which would over dilute the higgsino abundance. Therefore, we must search for a candidate for $X$ which decays more quickly.

One possibility is to consider unifying the origin of $\mu$ with the particle $X$. Specifically, consider charging $X$ and the $H_{u,d}$ chiral multiplets under a $U(1)_{PQ}$ and using a Kim-Nilles superpotential ~\cite{Kim:1983dt} of the form
\begin{align}
    W \supset \lambda\,\frac{X^n H_u H_d}{\Mpl^{\,n-1}},\qquad \mu = \lambda\,\frac{\expval{X}^n}{\Mpl^{\,n-1}},
\end{align}
with $n$ fixed by PQ charge assignments. Related PQ-based interpretations of the LZ event have been
discussed in Refs.~\cite{Yin:2026jnn,Visinelli:2026kgt}.

We decompose the scalar component as
\begin{align}
    X = \expval{X}\left(1 + \frac{s}{\sqrt{2}\expval{X}}\right)e^{ia/\sqrt{2}\expval{X}}
\end{align}
The potential for $s$ only arises after SUSY breaking. We assume the radial component, $s$, starts with Plankian field displacement after inflation and that the minimum of the potential gives the field a mass $m_s$ such that the field begins to oscillate about its minimum well before the higgsino freeze-out temperature subsequently dominates the energy density of the universe.

We identify the coupling of $s$ to the higgs fields through the replacement $\mu \to \mu(X)$. We assume $m_s <2 m_\chi$ and that the axino is too heavy to decay into as could be the case in gravity mediated SUSY breaking models. Then identifying $\mu \approx m_\chi$ we estimate the summed visible
decay width as~\cite{Co:2017orl}
\begin{align}
    \Gamma_{s\to {\rm SM}} \approx \frac{n^2\,m_\chi^4}{4\pi \expval{X}^2 m_s}
\end{align}
For $n=3$, $\expval{X} \approx 2\lambda^{-1/3}\times 10^{14}\, (m_\chi/10^6\GeV)^{1/3} \GeV$ such that $s$ decays to reheat the SM plasma with
\begin{align} \label{eq:TRH}
    \TRH \approx 3\TeV\,\lambda^{1/3}\left(\frac{m_\chi}{10^6\GeV}\right)^{7/6}\left(\frac{m_s}{2 m_\chi}\right)^{-1/2}.
\end{align}
We get the correct $\TRH$ of Eq.~\eqref{eq:TRH} to dilute higgsino DM to the correct abundance for
\begin{align}
    \lambda \approx 0.1\,(m_s/2m_\chi)^{3/2}(m_\chi/10^6\GeV)^{-5/2}.
\end{align}
From this we see the higgsino mass range of interest, $m_\chi = (10^5,\, 10^6)\GeV$, is allowed for $2m_h < m_s < 2m_\chi$. 

The field $s$ also decays to axions through its kinetic term, with $\Gamma_{s\to aa} = m_s^3/64\pi\expval{X}^2$, such that ${\rm BR}(s\to aa) \approx m_s^4/16n^2m_\chi^4$ and the axions contribute $\Delta N_{\rm eff} \approx 0.1\,(m_s/1.5\,m_\chi)^4$. Bounds from $\Delta N_{\rm eff} $ are avoided for $m_s <1.5\,m_\chi$ and this contribution to $\Delta N_{\rm eff} $ is further suppressed for smaller $m_s/m_\chi$.

One less appealing aspect of this proof of concept model is that we must tune the axion misalignment angle to $\theta\lesssim 0.05$ to avoid axions making up more that $10\%$ of DM due to the large value of $\expval{X}$.

\section{Conclusion} \label{sec:Conclusion}
If the high recoil energy event at LZ is in fact from inelastic scattering with higgsino DM, the results of \cite{Rodd:2026tyn} and \cite{Pospelov:2026ewn} suggest a strong tension with the interpretation that is comes for a collision with a thermal higgsino with $m_\chi = 1.1\TeV$. However, given that the higgsino is compelling in its own right and that the required mass $\delta$ already suggests the existence of a very high scale of $10^7$~GeV, it is worth considering the possibility that higgsino DM may also close to this scale. This allows reasonable consistency with both the LZ results and solar capture bounds. 

This comes at the cost of abandoning the WIMP miracle and requiring modifications to the early cosmological history to dilute the otherwise overly abundant heavy higgsinos. In this paper we give one proof of concept model where this can be achieved, but other methods may exist to achieve the same goal.

Additionally, the fit demonstrates an interesting point. While the light global minimum of the likelihood fit for $m_\chi = (200,\,400)\GeV$ is excluded by the solar capture bounds (specific to particles which annihilate to final states with neutrinos), it performs well at explaining the absence of events in the high energy sideband. While nuclear recoil distributions depend strongly on the type of interactions involved, the general feature that lighter DM aids in this effort may suggest the possibility that even if this event comes from a different DM candidate than a Higgsino, it still may be searched for at future colliders.

\begin{acknowledgments}
\noindent 
I would like to thank Matthew Reece and Tracy Slatyer for very valuable discussions. This material is based upon work supported by the U.S. Department of Energy, Office of Science, Office of High Energy Physics of U.S. Department of Energy under grant Contract Number  DE-SC0012567."(High Energy Theory research) and Simons Foundation Investigation Awards 929255 and 929241.
\end{acknowledgments}

\bibliography{bibliography}

\end{document}